\documentclass[runningheads]{llncs}
\usepackage{graphicx}
\usepackage{booktabs}
\usepackage{hyperref}
\usepackage{url}
\usepackage{caption}
\usepackage{subcaption}

\usepackage{algorithm}
\usepackage{algorithmic}
\usepackage{amsmath}
\usepackage{wrapfig}
\begin{document}
\title{Description Generation using Variational Auto-Encoders for precursor microRNA}
\titlerunning{Description Generation for pre-miRNA}
\author{Marko Petkovi\'c\inst{1}\orcidID{0009-0000-4918-6027} \and
Vlado Menkovski\inst{1}\orcidID{0000-0001-5262-0605}}
\authorrunning{M. Petkovi\'{c} \and V. Menkovski}
\institute{Eindhoven University of Technology, The Netherlands
\email{\{m.petkovic1,v.menkovski\}@tue.nl}}
\maketitle              % typeset the header of the contribution
\begin{abstract}
Micro RNAs (miRNA) are a type of non-coding RNA, which are involved in gene regulation and can be associated with diseases such as cancer, cardiovascular and neurological diseases. As such, identifying the entire genome of miRNA can be of great relevance. Since experimental methods for novel precursor miRNA (pre-miRNA) detection are complex and expensive, computational detection using ML could be useful. Existing ML methods are often complex black boxes, which do not create an interpretable structural description of pre-miRNA. In this paper, we propose a novel framework, which makes use of generative modeling through Variational Auto-Encoders to uncover the generative factors of pre-miRNA. After training the VAE, the pre-miRNA description is developed using a decision tree on the lower dimensional latent space. Applying the framework to miRNA classification, we obtain a high reconstruction and classification performance, while also developing an accurate miRNA description.
 
\end{abstract}
\section{Introduction}

% In recent years, generative models such as Variational Auto-Encoders (VAEs) and Generative Adversarial Networks (GANs) have been successfully applied to a wide variety of tasks, like BLABLABLA CITE. Typically, VAEs and GANs generate data by transforming a low-dimensional latent space to higher dimensional data. The low-dimensional latent space in these models is usually entangled, meaning that one latent variable can encode multiple aspects of the generative process. To tackle this problem, different disentanglement techniques have been proposed \cite{higgins2016beta,chen2018isolating,ilse2020diva}, where the model is forced to learn a latent representation in which each variable only encodes one aspect of the generative process. However, it is not inherently clear which latent variable corresponds to which property of the data, and how these properties could be used to describe the class of a datapoint. 

In living organisms, DNA encodes all the information used by the organism to develop and survive. To make use of the information encoded in the DNA, it needs to be translated to RNA. One type of RNA is messenger RNA (mRNA), which is used to create proteins. Sometimes, depending on the cell type, it is desired that certain mRNA is not expressed, which is where micro RNA (miRNA) comes into play, by silencing and inhibiting the expression of certain mRNAs. MiRNAs can silence mRNAs by binding to them, following which the mRNA is degraded. As such, the definition of what constitutes a miRNA is functional, since it needs to carry out this task. Depending on the shape of the miRNA, it can be used to silence (multiple) different mRNAs. Since over- and under-expression of miRNA can lead to various diseases, it is important to understand the genome of miRNA better. 

\begin{figure}[h]
    \centering
    \includegraphics[width=.9\linewidth]{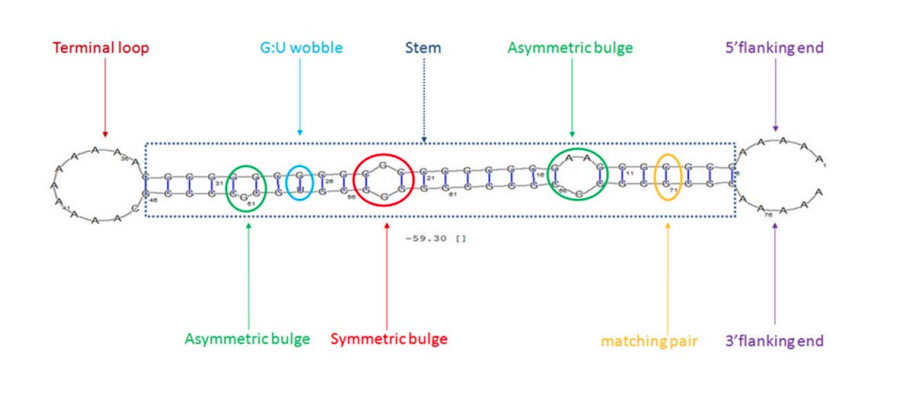}
    \caption{Artificial pre-miRNA strand \cite{allmer2014computational}, labeled with different properties which can be present in (non) pre-miRNA.}
    \label{fig:premirna}
\end{figure}

It is estimated that there are around 2300 types of human miRNA \cite{alles2019estimate}, while only around 700 are documented \cite{kozomara2019mirbase}. Finding new miRNAs often requires complex lab conditions \cite{ng2007novo,saccar2014machine}, where potential new miRNA needs to be differentiated from other RNA. When identifying novel miRNAs, precursor miRNA (pre-miRNA) is typically used. This is an RNA molecule that is transformed into miRNA in the cell. Pre-miRNA is folded in a hairpin and contains more nucleotides and thus has more features (see Figure \ref{fig:premirna}). As a result, it is easier to differentiate from other RNA. While some features, like the presence of a terminal loop and a large fraction of base pairs (C-G/A-U) in the stem, are prevalent in pre-miRNA, they are not enough to differentiate it from other RNA molecules and hence are not sufficient to form a structural description of pre-miRNA. 

A number of data-driven methods have been developed for identifying novel pre-miRNA. Some of the pre-miRNA detection methods are based on traditional ML methods using engineered features \cite{saccar2017performance}, such as Random Forests (RFs) \cite{jiang2007mipred} and Support Vector Machines (SVMs) \cite{batuwita2009micropred,ding2010mirensvm,jiang2007mipred,ng2007novo,xue2005classification}, where only limited interpretability is possible through feature importance. In addition, several Deep Learning methods have been proposed, which rely on Convolutional Neural Networks (CNNs) \cite{Cordero840579,do2018precursor,tasdelen2021hybrid,zheng2019nucleotide} and Recurrent Neural Networks (RNNs) \cite{park2017deep,tasdelen2021hybrid}. While these methods are able to achieve state-of-the-art accuracy on the miRNA classification task, interpretation is only possible through methods using activation maps, or techniques such as concept whitening \cite{brandtconcept}. As such, these black-box models do not directly allow us to form a description of pre-miRNA that would in turn allow us to understand better these molecules and their roles in the cell. 

In recent years, generative models such as Variational Auto-Encoders (VAEs) and Generative Adversarial Networks (GANs) have been successfully applied to a wide variety of tasks, such as protein \cite{ingraham2019generative}, drug \cite{cheng2021molecular} and DNA design \cite{killoran2017generating}. Typically, VAEs and GANs generate data by mapping a low-dimensional latent space to high-dimensional data. The low-dimensional latent space in these models is usually entangled, meaning that one latent variable can encode multiple aspects of the generative process. To tackle this problem, different disentanglement techniques have been proposed \cite{chen2018isolating,higgins2016beta,ilse2020diva}, where the loss function or model architecture forces the model to learn a latent representation in which each variable only encodes one aspect of the generative process. However, it is not inherently clear which latent variable corresponds to which property of the data, and how these properties could be used to describe the class of a datapoint.
% as well as Deep Learning \cite{Cordero840579,do2018precursor,park2017deep,tasdelen2021hybrid,zheng2019nucleotide}. These methods typically make use of precursor miRNA (pre-miRNA), which is folded in a hairpin shape and has more discriminative features. While these methods achieve a high performance, due to their complexity, they often provide very limited interpretability as to why an RNA strand is or is not a miRNA.  

In this work, we develop a method for developing pre-miRNA descriptions using the lower dimensional latent space of a latent variable model. Our contributions are two-fold: (i) we propose a novel framework based on VAEs and Decision Trees (DTs), which can be used to develop interpretable descriptions of different object classes $y$ being modeled, in terms of features $f$ and (ii) apply this framework to the domain of precursor microRNA, where we demonstrate that it obtains biologically sound structural descriptions.

% \section{Related Work}
% In the basic version of the VAE \cite{kingma2013auto}, the latent space typically becomes organized, with datapoints with similar features being close together in terms of their latent embeddings. However, a single latent factor might encode (parts of) multiple generative factors. In $\beta$-VAE \cite{higgins2016beta}, a penalty term ($\beta$) was added to the KL term of the ELBO, which forces the model to learn independent latent variables. As a result, each latent variable learns to encode a separate feature. To figure out which variable encodes which feature, inspection of reconstruction is required. A similar approach was proposed in $\beta$-TCVAE \cite{chen2018isolating}, where the KL-term is further decomposed into a mutual information, total correlation and dimension-wise KL term. Furthermore, each term in the decomposition is assigned its own penalty term. A different approach to disentanglement was taken for the DIVA model \cite{ilse2020diva}, where three separate latent spaces are used, for the data domain, class and remaining variance. Each latent space has its own encoder, while the latent space for the domain and class in addition have conditional priors and auxiliary classifiers on top of the latent space. As a result, the model is forced to disentangle variation from the domain and class of the data. 

\section{Methods}
\subsection{Data}

\begin{figure}[h]
    \centering
    \includegraphics[width=.9\linewidth]{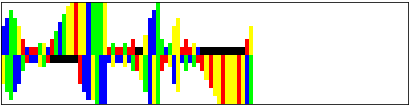}
    \caption{Example of RNA image encoding.}
    \label{fig:rnaimage}
\end{figure}

We use the \textit{modmiRBase} dataset \cite{Cordero840579}, which contains 49602 RNAs from various organisms, and is balanced in terms of class. The pre-miRNA sequences in the dataset were taken from \url{mirbase.org} \cite{griffiths2004microrna} and \url{mirgenedb.org} \cite{mirGenedb}, while the non pre-miRNA sequences were taken from \cite{gudys2013huntmi,ng2007novo,saccar2017performance,wei2013improved}. The RNA sequences are folded using the RNAFold algorithm \cite{hofacker1994fast} , following which they are encoded as images of 100 by 25 pixels \cite{Cordero840579}. In the image encoding, each nucleotide is represented as a colored bar (A: blue, C: yellow, G: green and U: red), while gaps (black) are introduced for nucleotides without an opposing nucleotide. The length of each bar inversely depends on the bond strength between nucleotides, where bonds between base pairs are considered strong, and bonds between other pairs or gaps are considered weak.  The length of each bar additionally increases with multiple subsequent weak bonds (gaps always have a length of 2 pixels). As such, the image encoding resembles the physical shape of the RNA. An example of an image encoding can be found in Figure \ref{fig:rnaimage}. 

Since the shape plays an important role in the process of silencing mRNAs, we also encode the bond strength between nucleotides. This is done in vector $m$ of size 100, where a strong bond is assigned a value of 1, while a weak bond (or no bond) is assigned a value of 0.

For each datapoint, we also calculated multiple features, using the algorithms from \cite{brandtconcept}. Several of the features can be found in Figure \ref{fig:premirna}, such as the presence of a terminal loop, the size of the terminal loop, the length of the stem etc. These features are not used directly in our VAE, but are used in the decision tree algorithm instead.

\subsection{Model}
\begin{figure}[h]
    \centering
    \includegraphics[width=0.9\textwidth]{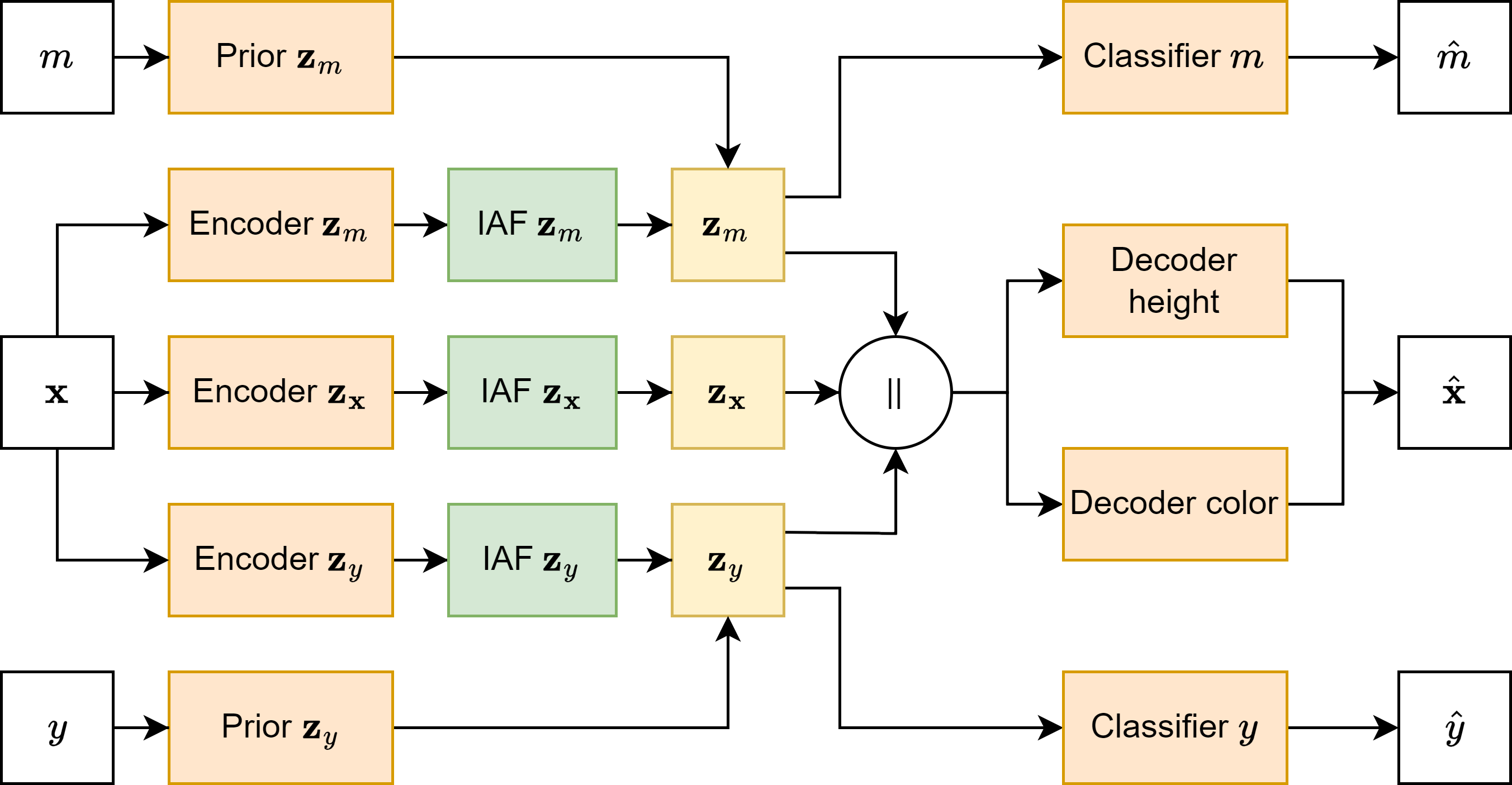}
    \caption{Full model architecture. $\|$ represents concatenation.}
    \label{fig:diva}
\end{figure}

\begin{figure}[h!]
    \centering
    \begin{subfigure}[b]{0.9\textwidth}
         \centering
         \includegraphics[width=\textwidth]{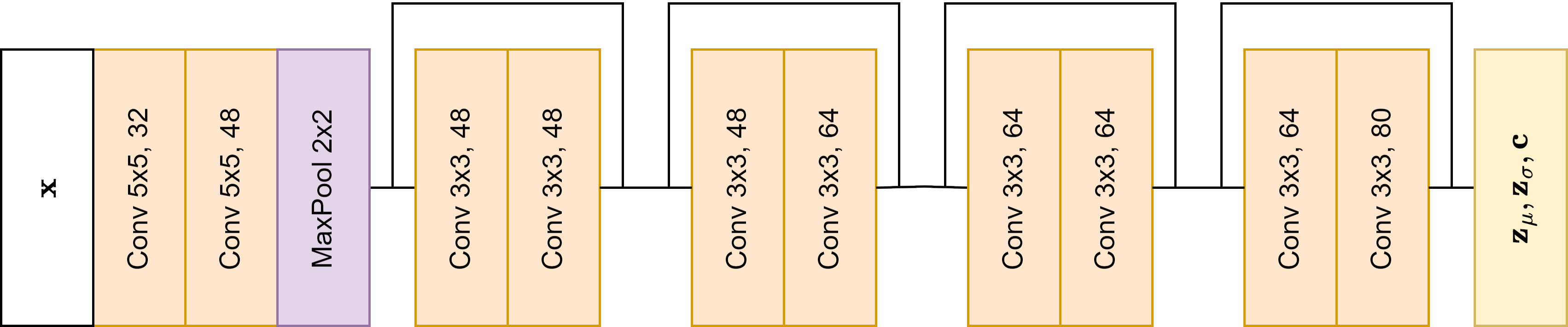}
         \caption{Encoder architecture. Sampling from the final layer is followed by IAF to obtain the latent space.}
         \label{fig:encoder}
     \end{subfigure}\\
     \vspace{2mm}
    \begin{subfigure}[b]{0.45\textwidth}
         \centering
         \includegraphics[width=\textwidth]{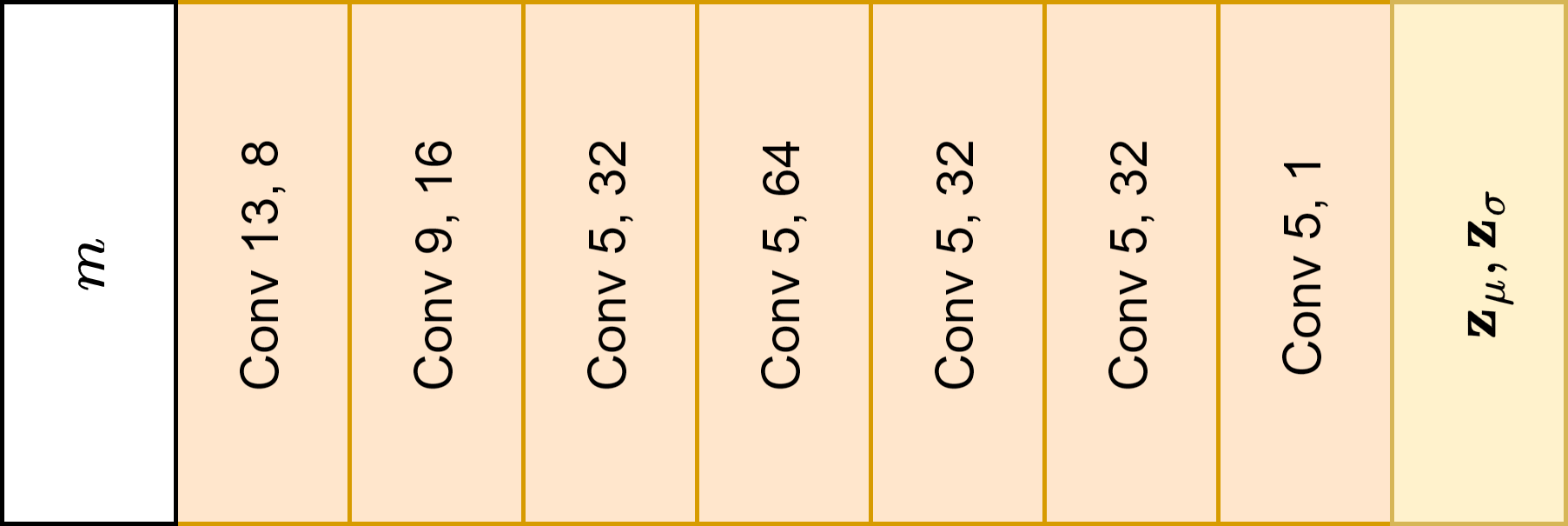}
         \caption{Prior ($\mathbf{z}_m$) architecture.}
         \label{fig:prior}
     \end{subfigure}
     \hfill
     \begin{subfigure}[b]{0.5\textwidth}
         \centering
         \includegraphics[width=\textwidth]{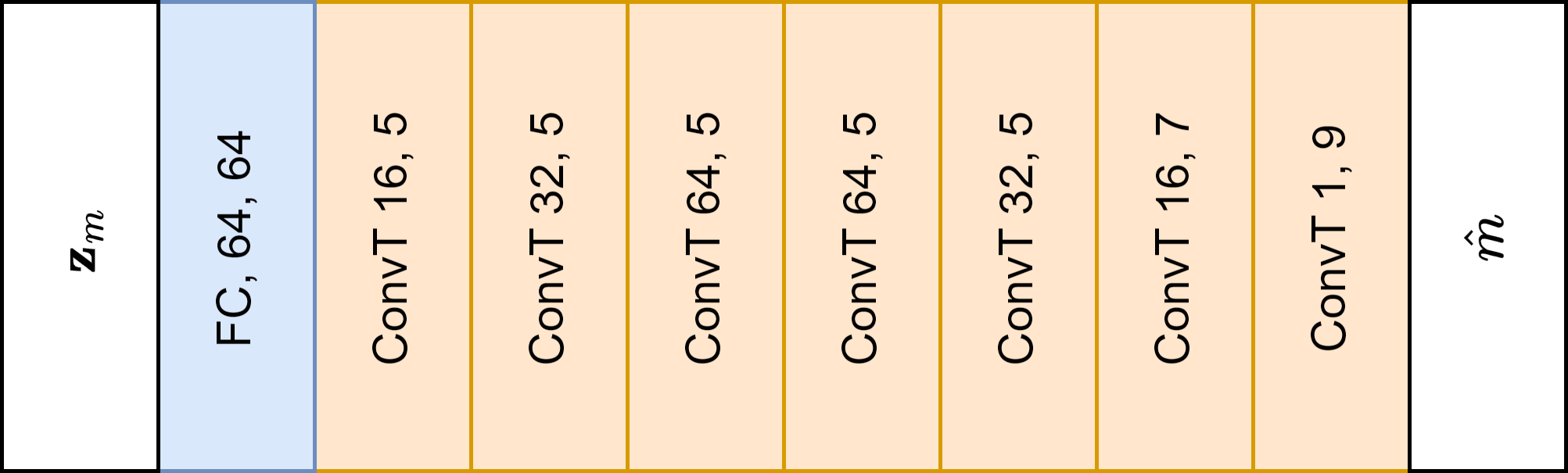}
         \caption{Classifier for $m$ architecture.}
         \label{fig:cls}
     \end{subfigure}\\
     \vspace{2mm}
     \begin{subfigure}[b]{0.9\textwidth}
     \centering
     \includegraphics[width=0.65\textwidth]{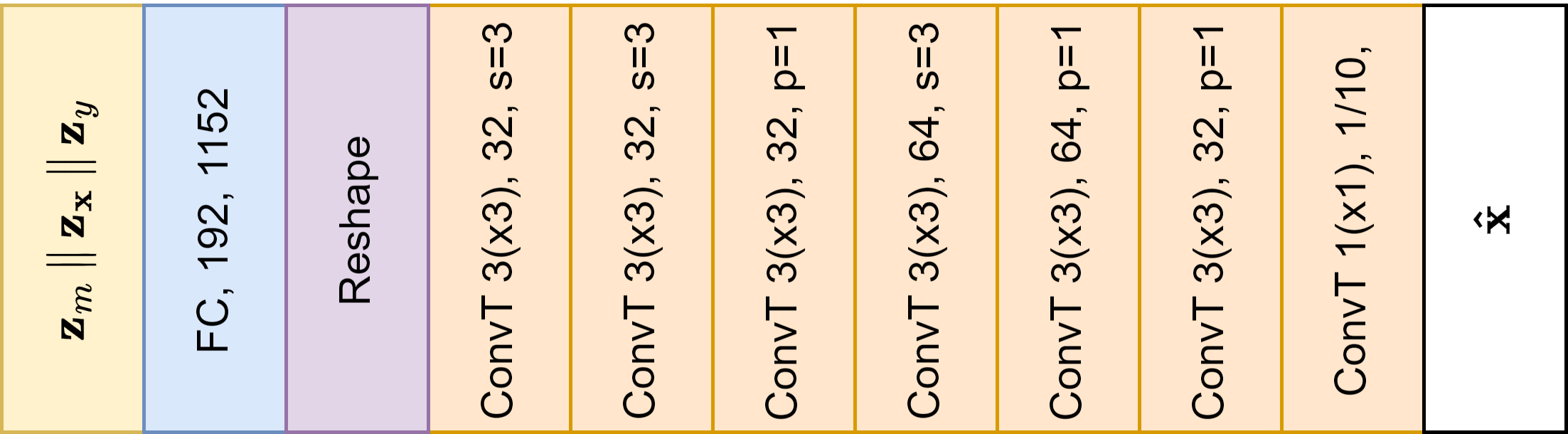}
     \caption{Decoder for bar height/color. Both decoders are preceded by the same fully connected layer and reshape operation. The height decoder uses 2D transposed convolutions, while the decoder for color first sums over the height dimension and then processes uses 1D transposed convolutions. For the final layer, the color decoder gives the probability for each color (top and bottom row, 5 colors in each bar). Outputs of the two decoders are combined to obtain $\mathbf{\hat{x}}$}
     \label{fig:decoder}
     \end{subfigure}
    \caption{Architecture of different model components. First number following the layer name indicates the kernel size, the second number the amount of filters. Unless mentioned otherwise, each convolutional layer uses a stride (s) of 1. Convolutional layers in the encoder use same padding (p), while other layers use valid padding unless mentioned otherwise.}
    \label{fig:modelcomps}
    \vspace{-5mm}
\end{figure}

To model the (non) pre-miRNA, we made use of the DIVA model \cite{ilse2020diva}, with a latent space for bond strength ($\mathbf{z_m}$), class ($\mathbf{z}_y$) and remaining variance ($\mathbf{z}_x$). Each latent space has a size of 64. For each latent space, a separate encoder based on ResNet \cite{he2016deep} was used, with an overview of the architecture in Figure \ref{fig:encoder}. Furthermore, $\mathbf{z_m}$ and $\mathbf{z}_y$ have conditional priors, whereas $\mathbf{z}_x$ has a standard normal prior. The prior for $\mathbf{z_m}$ consists of a 1D convolutional network (Figure \ref{fig:prior}), followed by a fully connected layer, while the prior of $\mathbf{z}_y$ consists of two fully connected layers. In addition, auxiliary classifiers were used to further enhance the disentanglement. The auxiliary classifier for $\mathbf{m}$ consists of a fully connected layer follow by 1D deconvolutional layers (Figure \ref{fig:cls}), while the auxiliary classifier for $y$ consists of two fully connected layers. Since the shape of the miRNA place an important role in the mRNA silencing process, we made  use of an additional auxiliary classifier for $y$ on $\mathbf{z_m}$. Finally, to organize the latent space better, Inverse Autoregressive Flow (IAF) \cite{kingma2016improved} was applied on all 3 latent spaces. Each IAF consists of 8 blocks of 2 Masked Autoencoder for Distribution Estimation (MADE) \cite{germain2015made} layers, which have a context size of 32 (obtained from the encoder), as well as a hidden size of 1080. 

Since our data is highly structured, both the encoders and decoders with the data complexity in mind. In the input representation, we one-hot encoded the 5 possible different colors, to ensure that distances between different colors are always the same, which results in an image with 5 channels. Furthermore, using a regular deconvolutional image decoder could yield invalid reconstructions, as respecting the structure of the data would not be possible. To limit the amount of invalid images, we separately decode the height and color of each bar. After both height and color distributions are decoded, we combine the two distributions to obtain a distribution over both height and color, which is used to calculate the reconstruction loss. The decoder of the bar height is a 2D deconvolutional decoder. It outputs the probability for the different possible bar heights, after softmax activation. To obtain the actual bars, the probability distribution is multiplied with the different bar heights, similar to the approach in StampNet \cite{visser2019stampnet}. The decoder for the bar color is a 1D deconvolutional decoder, and outputs the color probabilities of the bars in the bottom and top row. Finally, the heights and colors are multiplied to obtain the reconstruction. When sampling from the decoder, a discrete value should be sampled from both the height and color distributions. The architecture of both decoders can be found in Figure \ref{fig:decoder}.

An overview of the full model architecture can be found in Figure \ref{fig:diva}. In the model, each layer is followed by batch normalization, and uses the ELU activation function (unless mentioned otherwise). For training the model, the loss functions in Equation \ref{eq:diva2} was used. In Equation \ref{eq:diva1}, the first term represents the reconstruction loss, while the remaining terms represent the KL divergence between the encoded prior ($q_\phi$) and the (learned) prior ($p/p_\theta$). The penalty terms to each KL divergence are represented by $\beta$. In Equation \ref{eq:diva2}, each additional term represents the loss of the different auxiliary classifiers ($q_\omega$), with $\alpha$ being the penalty term for each classifier. The dataset was split into a training set containing 34721 samples and a test set containing 14881 samples, with both datasets being balanced. For training we set each $\beta$ parameter to 0.5. Both $\alpha_{y1}$ and $\alpha_{y2}$ were set to 12, while $\alpha_m$ was set to 1. The model was trained using the Adam optimizer \cite{kingma2014adam}, a batch size of 64 and a learning rate of 0.0005 until convergence. 

\begin{equation}
\begin{split}
        \mathcal{L}_s(m,\mathbf{x},y) = \text{ }  &E_{q_{\phi_m}(\mathbf{z}_m|\mathbf{x}), q_{\phi_x}(\mathbf{z_x}|\mathbf{x}), q_{\phi_y}(\mathbf{z}_y|\mathbf{x}) }   [\log(p_\theta(\mathbf{x} \, | \, \mathbf{z}_y, \mathbf{z_x},\mathbf{z}_d))]\\
    &- \beta_m * D_{KL}(q_{\phi_m}(\mathbf{z}_m \, | \, \mathbf{x}) \; || \; p_{\theta_m}(\mathbf{z}_m \, | \, m)) \\
    &- \beta_x * D_{KL}(q_{\phi_x}(\mathbf{z_x} \, | \, \mathbf{x}) \; || \; p(\mathbf{z}_x)) \\
    &- \beta_y * D_{KL}(q_{\phi_y}(\mathbf{z}_y \, | \, \mathbf{x}) \;  || \; p_{\theta_y}(\mathbf{z}_y \, | \, y)) \\
    \label{eq:diva1}
\end{split}
\end{equation}
\vspace{-5mm}
\begin{equation}
\begin{split}
    \mathcal{F}_\text{DIVA}(m,\mathbf{x},y) &= \mathcal{L}_s(m,\mathbf{x},y)\\
    &+ \alpha_{y_{1}} E_{q_{\phi_{y}(\mathbf{z}_y|\mathbf{x})}}[\log q_{\omega_{y}}(y \, | \, \mathbf{z}_y)]\\
    &+ \alpha_{y_{2}} E_{q_{\phi_{y}(\mathbf{z}_m|\mathbf{x})}}[\log q_{\omega_{y}}(y \, | \, \mathbf{z}_m)]\\
    &+ \alpha_m E_{q_{\phi_{m}(\mathbf{z}_m|\mathbf{x})}}[\log q_{\omega_{m}}(m \, | \, \mathbf{z}_m)]
    \label{eq:diva2}
\end{split}
\end{equation}

To verify that the model architecture is optimal, we performed an ablation studies where we compared the aforementioned model with a standard(/$\beta$) VAE using a decoder consisting of fully connected layers (VAE), a $\beta$-VAE with IAF applied on the latent space ($\beta$-IAF-VAE), as well as with a deconvolutional decoder and IAF (DC-$\beta$-IAF-VAE). For all models, $\beta$ was set to 0.5. Code for all models and experiments can be found on \url{github.com/marko-petkovic/mirna}.

\subsection{Decision Tree}
\begin{algorithm}[tb]
\caption{\textbf{MakeSplit}($\mathbf{z}, y, f$) }\label{alg:1}
    \begin{algorithmic}[1]
    \REQUIRE $max\_depth, min\_samples, min\_acc$
        \STATE $\text{Node} \gets Node(\mathbf{z}, y, concepts)$
        \IF{current depth $\geq max\_depth$ \OR $len(y) < min\_samples$ \OR Node is pure}
        \RETURN Node
        \ENDIF
        \STATE gain $\gets$ 0, cls $\gets$ None
        \FOR{$feature$ in $f$}
        \STATE  train SVM $(X=\mathbf{z}, y = feature)$
        \STATE  split $\mathbf{z}$ based on classifier predictions in $\mathbf{z_0}$, $\mathbf{z_1} $
        \STATE calculate information gain based on split
        \IF{information gain $>$ gain \AND SVM accuracy $> min\_acc$}
        \STATE $\text{gain} \gets \text{information gain}$
        \STATE cls $\gets$ SVM
        \ENDIF
        \ENDFOR
        \STATE Split data into $\mathbf{z_0}$, $\mathbf{z_1}$, $concepts_0$, $concepts_1$, $y_0$, $y_1$ according to cls
        \STATE Node.left\_child $\gets$ \textbf{MakeSplit}($\mathbf{z_0}$,$y_0$,$f_0$,$ depth$)
        \STATE Node.right\_child $\gets$ \textbf{MakeSplit}($\mathbf{z_1}$,$y_1$,$f_1$,$depth$)
        \RETURN Node

    \end{algorithmic}
\end{algorithm}

To develop a description of pre-miRNA, we create a decision tree algorithm on top of the $\mathbf{z}_m$ latent space, as this latent space contains both information regarding the shape and class of RNA. Since this latent space is organized, various features of the RNA should be linearly separable within this space. As such, a linear SVM can be trained to partition the space based on a feature. 

Compared to a normal decision tree algorithm, such as  CART \cite{breiman2017classification} or C4.5 \cite{quinlan2014c4}, our algorithm makes splits based on the learned latent representation rather than the features ($f$) themselves. To create a split using our decision tree algorithm, for each feature, we train an SVM to partition the latent space ($\mathbf{z}_m$), with the feature being the target variable. In case a feature is continuous, we make multiple binary features out of it, where the feature is divided according to different thresholds. Alternatively, one could train a regression model rather than a classifier, and decide the threshold for a binary feature afterwards. For each split, we assess whether the performance of the SVM is above a threshold, and only keep the SVMs with a high enough performance. Then, to decide which split to make, we asses which split yields the highest information gain based on the class of the RNA. After training a full decision tree, we can assess which descriptions of pre-miRNA the model developed, by following paths down the tree which lead to a pre-miRNA classification. The full algorithm for making splits for the DT can be found in Algorithm \ref{alg:1}.

To perform predict whether a new datapoint is a pre-miRNA, it should first be encoded by the encoder for $\mathbf{z}_m$. Then, the decision tree should be followed, where it is checked on which side of the split by the SVM the point lies. 

After training the model, we trained a decision tree with $max\_depth = 5$, $min\_samples = 10$ and $min\_acc = 0.8$ using the training set and features $f$ from \cite{brandtconcept}. Each of the continuous features were partitioned into multiple binary features.

\section{Results}
\subsection{Performance}

In Table \ref{tab:recstats}, the reconstruction performance of our model and the different ablations can be found. The reconstruction performance was calculated based on three different metrics, the MAE on the entire image, the MAE based on nucleotide color and the MAE in terms of length of the nucleotides. As can be seen, our model achieved the highest performance both in terms of all three metrics.

\begin{table}[b]
\centering
\caption{Reconstruction statistics for different model types. Bold statistics represent lowest obtained errors.} 
\label{tab:recstats}
\begin{tabular}{lccc}
\toprule
Model Name &     MAE &  MAE nucleotide &  MAE length \\
\midrule
VAE                &  0.136 &          0.305 &      0.784 \\
$\beta$-VAE        &  0.055 &          0.092 &      0.478 \\
$\beta$-IAF-VAE    &  0.053 &          0.082 &      0.444 \\
DC-$\beta$-IAF-VAE &  0.009 &          0.015 &      0.067 \\
DC-IAF-DIVA        &  \textbf{0.007} &         \textbf{0.012} &      \textbf{0.057} \\
\bottomrule
\end{tabular}
\end{table}
\begin{figure}[h]
    \centering
    \includegraphics[width=.9\textwidth]{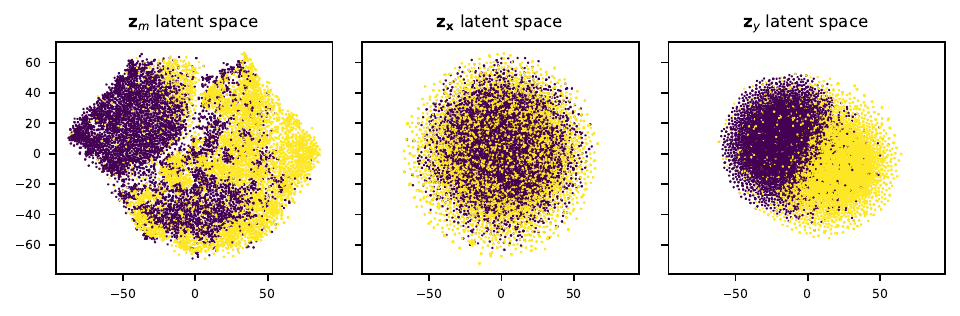}
    \caption{Latent spaces of DIVA model after dimensionality reduction using t-SNE. Yellow dots represent pre-miRNA.}
    \label{fig:diva_ls}
\end{figure}

\begin{figure}[h]
    \centering
    \includegraphics[width=.9\textwidth]{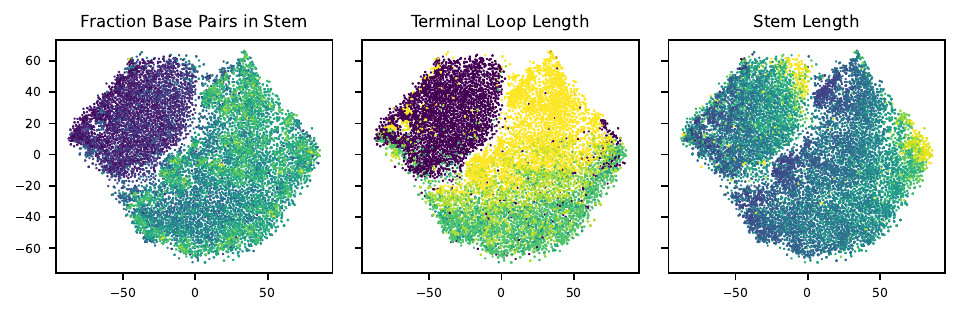}
    \caption{$\mathbf{z}_m$ latent space colored according to the fraction base pairs in stem, terminal loop length and stem length. Lighter colors indicate higher values for each variable.}
    \label{fig:diva_zm}
\end{figure}

To verify that the model has learned an organized latent space, we have visualized the latent spaces using t-SNE \cite{van2008visualizing} dimensionality reduction. As can be seen in Figure \ref{fig:diva_ls}, the disentanglement of the DIVA model indeed appears to be successful, as there seems to be a clear separation between classes in $\mathbf{z}_m$ and $\mathbf{z}_y$, while this is not present in $\mathbf{z_x}$. Taking a further look at the distribution of the fraction of base pairs in the stem, terminal loop length and the stem length over the $\mathbf{z}_m$ latent space (Figure \ref{fig:diva_zm}), we see that the latent space is well organized. Therefore, it can be used by our decision tree algorithm to develop a pre-miRNA description.

\subsection{Conditional Generation}

In addition to inspecting the latent space embeddings, we also qualitatively verify the learned representation of $\mathbf{z}_m$. Here, we first encode a selected RNA image to obtain $\mathbf{z}_m$, $\mathbf{z}_y$ and $\mathbf{z_x}$. Following this, the bond strength $m$ of the sample is modified, and passed through the conditional prior, to obtain $\mathbf{z}_m'$. Afterwards, we reconstruct the RNA based on $\mathbf{z}_m'$, $\mathbf{z}_y$ and $\mathbf{z}_x$. Finally, we verify whether the newly generated RNA image indeed follows the provided bond strength.

\begin{figure}[tb]
    \centering
    \includegraphics[width=0.9\textwidth]{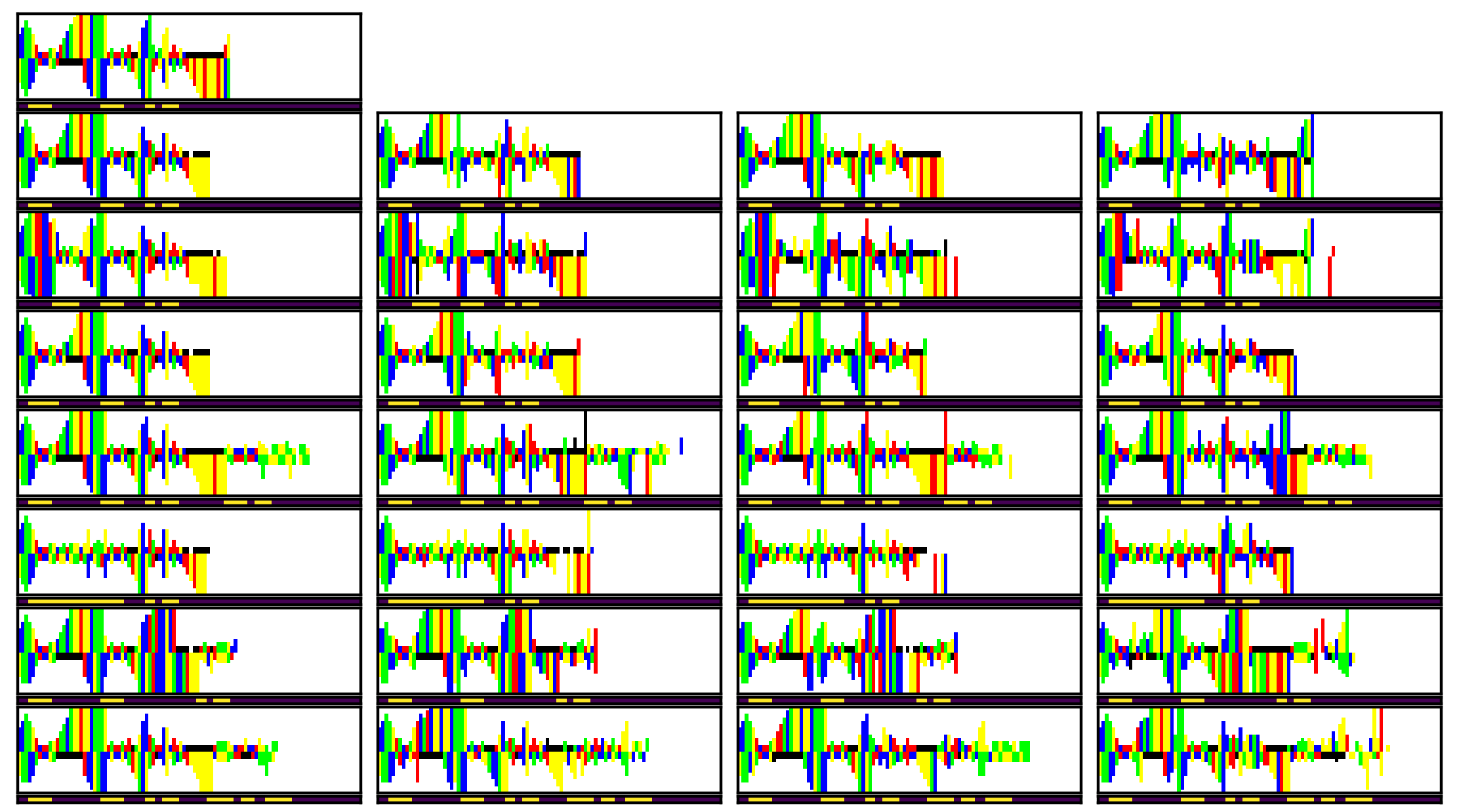}
    \caption{Generated RNA images, with $\mathbf{z}_m$ sampled from the prior of various $m$. In each row, a different $m$ was used. In the second row, the same $m$ as in the original image was used.}
    \label{fig:priorchange}
\end{figure}

In Figure \ref{fig:priorchange}, we can find the results of the conditional generation experiment. In the top left, the original image can be found, with the bond strength $m$ displayed below the picture. In each column, a different sample from the same conditional prior is reconstructed, while for each row a unique $m$ was used. Overall, we see that there are a few (minor) errors in reconstruction, but in general, the generated images follow the provided bond strength $m$. As such, we can conclude that the learned latent space is well organized both in terms of features and bond strength.

\subsection{Decision Tree}

\begin{wrapfigure}{r}{0.6\textwidth}
\vspace{-20mm}
\begin{center}
\includegraphics[width=0.55\textwidth]{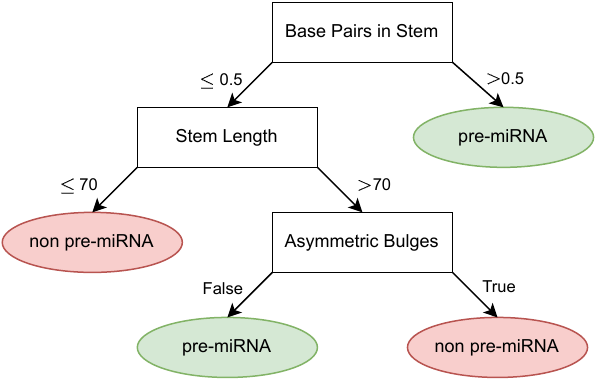}
\end{center}
\caption{Learned pre-miRNA description.}
\vspace{-8mm}
\label{fig:dt}
\end{wrapfigure}

The obtained description from the decision tree can be found in Figure \ref{fig:dt}. Overall, the model achieved an accuracy of 0.912, a sensitivity of 0.923 and a specificity of 0.955. The concept whitening method achieved an accuracy of 0.924 on the same dataset, while offering less interpretation compared to the decision tree.

% \begin{figure}[hb]
%     \centering
%     \includegraphics[width=0.6\textwidth]{figures/dt.pdf}
%     \caption{Learned pre-miRNA description}
%     \label{fig:dt}
% \end{figure}

In addition to the high classification performance, the decision tree learned biologically sound classification rules. The first split in the decision tree is made based on the amount of base pairs in the stem, where samples with a high amount of base pairs is considered to be an pre-miRNA, which is also supported by literature \cite{allmer2014computational}. Similarly, short RNA sequences are also considered unlikely to be pre-miRNA \cite{allmer2014computational}, which can be seen in the second split of the tree. The final feature used by the tree to discriminate pre-miRNA from other RNA is the absence of large asymmetric bulges, which is also a known property of pre-miRNA \cite{kang2015computational}. In general, given the supplied features, we see that the decision tree is able to generate a biologically valid description of pre-miRNA.  

\section{Discussion}
In this work, we have proposed a new method for latent space interpretability in VAEs, and applied it to pre-miRNA. The trained VAE for pre-miRNA generation has shown excellent performance, and was therefore suitable for application of our method. Finally, we have shown that the developed description matches existing knowledge of pre-miRNAs, suggesting that both the VAE and our interpretability method perform well. 
\newpage
\bibliographystyle{splncs04}
\bibliography{mybibliography}

\end{document}